\documentclass[12pt]{iopart}
\eqnobysec
\usepackage{iopams} 
\usepackage{amsthm} 
\usepackage{graphicx}
\usepackage{cite}

\begin{document}

\title[Inconsistency of the Bohm-Gadella theory with quantum mechanics]{On 
the inconsistency of the Bohm-Gadella theory with quantum mechanics}

\author{Rafael de la Madrid}
\address{Department of Physics, University of 
California at San Diego, La Jolla, CA 92093 \\
E-mail: {\texttt{rafa@physics.ucsd.edu}}}

\date{\small{(March 27, 2005)}}

\begin{abstract}
The Bohm-Gadella theory, sometimes referred to as the Time 
Asymmetric Quantum Theory of Scattering and Decay, is based on the Hardy 
axiom. The Hardy axiom asserts that the solutions 
of the Lippmann-Schwinger equation are functionals over spaces of 
Hardy functions. The preparation-registration arrow of time
provides the physical justification for the Hardy axiom. In this paper, it is 
shown that the Hardy axiom is incorrect, because the solutions of the 
Lippmann-Schwinger equation do not act on spaces of Hardy functions. It is 
also shown that the derivation of the preparation-registration arrow of time 
is flawed. Thus, Hardy functions neither appear when we solve the 
Lippmann-Schwinger equation nor they should appear. It is also shown that 
the Bohm-Gadella theory does
not rest on the same physical principles as quantum mechanics, and that it 
does not solve any problem that quantum mechanics cannot solve. The 
Bohm-Gadella theory must therefore be abandoned.


\end{abstract}


\section{Introduction}
\label{sec:introduction}

Recently, a new axiom for quantum mechanics, called the Hardy axiom,
has been introduced~\cite{FP03a}. This axiom is the bedrock of the
Bohm-Gadella theory, which is meant to be a ``{\it consistent and exact 
theory of quantum resonances and decay}''~\cite{IJTP03}. 

The Bohm-Gadella theory was initially formulated in the late 
1970s~\cite{LMP79,JPM80,JMP81}, and it was a creative attempt to accommodate 
states for nonrelativistic resonances, the ``Gamow 
vectors''~\cite{LMP79,JPM80,JMP81}. A byproduct of the Bohm-Gadella theory 
was that the time evolution of the ``Gamow vectors'' is given by a semigroup 
rather than by a group, expressing time asymmetry at the microscopic 
level. The Bohm-Gadella theory was subsequently refined 
in~\cite{JMP83a,JMP83b,JMP85}. With further improvements, it was summarized 
in~\cite{BG}. The single most important mathematical piece of the Bohm-Gadella 
theory is the rigged Hilbert space of Hardy class.

In 1994, the Bohm-Gadella theory received a boost from the 
preparation-registration arrow of time~\cite{PLA94,JMP95}. The 
preparation-registration arrow of time provides, by means of causality, the 
physical justification for the Hardy axiom and thus also for the Bohm-Gadella 
theory.

After several review articles, and after extending the theory to 
relativistic resonances~\cite{PLA00}, the assumptions
of the Bohm-Gadella theory have been explicitly formulated as a new axiom for
quantum mechanics, the Hardy axiom~\cite{FP03a}. The state of 
the art of the Bohm-Gadella theory can be found in review~\cite{IJTP03}. 

If we denote the Lippmann-Schwinger bras by $\langle ^{\pm}E|$, and the ``in'' 
and ``out'' wave functions by $\varphi ^{+}$ and $\psi ^-$, then the Hardy 
axiom states that 
$\langle ^{+}E|\varphi ^{+}\rangle \equiv \varphi ^{+}(E)$ and
$\langle ^{-}E|\psi ^{-}\rangle \equiv \psi ^{-}(E)$
are Hardy functions. By using the Hardy axiom, the two major 
achievements of the Bohm-Gadella theory, the ``Gamow vectors'' and time 
asymmetry, are readily obtained.

Although its goals are worth pursuing, the Bohm-Gadella theory suffers 
from at least two major problems. First, the Bohm-Gadella theory is purely 
formal, that is, nobody has found a potential to which such a theory 
applies. The way the Bohm-Gadella theory works is by 
{\it presupposing} that the solutions of the Lippmann-Schwinger equation are 
distributions acting on two spaces of Hardy functions. In the Bohm-Gadella
theory, the Lippmann-Schwinger equation is never solved, and one
gets along by means of the Hardy-function assumption. Second, and 
more important, the content of the Hardy axiom is not a matter of 
assumption, but a matter of proof. The properties of the solutions of the 
Lippmann-Schwinger equation cannot be simply assumed; 
rather, one must solve the Lippmann-Schwinger equation and derive the 
properties of its solutions.

The present paper is devoted to a critical examination of the Bohm-Gadella 
theory. We will show that the solutions of the Lippmann-Schwinger equation 
do not comply with the Hardy axiom, thereby showing that the Bohm-Gadella 
theory is inconsistent with standard quantum mechanics. In order to show this, 
we will use the
example of the spherical shell potential and the results of~\cite{LS1,LS2}, 
where that same potential was used to obtain the rigged Hilbert spaces that 
accommodate the Lippmann-Schwinger equation~\cite{LS1} and its analytic
continuation~\cite{LS2}.

As mentioned above, the justification for the Hardy axiom comes from causality,
by way of the preparation-registration arrow of time. Since the 
preparation-registration arrow of time seems so compelling, one might still
think that the Hardy axiom must be correct, and that there must be some
subtlety missing in our analysis. To dispel any remaining doubts, we 
will show that the derivation of the preparation-registration arrow of 
time is actually flawed. Thus, neither Hardy functions appear when we solve 
the Lippmann-Schwinger equation, nor there is a physical reason why they 
should.

In addition, we will see that not only the Hardy axiom but also other aspects 
of the Bohm-Gadella theory are inconsistent with quantum 
mechanics. Furthermore, the two major achievements of the Bohm-Gadella 
theory, the resonance states and time asymmetry, can be achieved within 
standard quantum mechanics. Thus, the Bohm-Gadella theory is not only 
inconsistent with quantum mechanics, but also unnecessary.

In Sec.~\ref{sec:HFreview}, we review the properties of Hardy functions. As 
well, we recall the precise statement of the Hardy axiom.

In Sec.~\ref{sec:alinfar}, we show that the behavior of $\varphi ^{+}(E)$ and 
$\psi ^{-}(E)$ does not comply with the Hardy axiom. More specifically, we 
will show that whereas the Hardy axiom implies that
the analytic continuation of $\varphi ^{+}(E)$ (respectively $\psi ^{-}(E)$) 
vanishes on the lower (respectively upper) infinite arc of the second sheet 
of the Riemann surface, the truly quantum mechanical $\varphi ^{+}(E)$ and
$\psi ^{-}(E)$ blow up exponentially in that infinite arc.

In Sec.~\ref{sec:PRAT}, we show that the derivation of the 
preparation-registration arrow of time is flawed.

In Sec.~\ref{sec:SpectrumH}, we argue that the semiboundedness of the
Hamiltonian seems to be the reason why the Hardy axiom does not apply
in quantum mechanics.

In Secs.~\ref{sec:GSdBSss} and~\ref{sec:GSdBS}, we compare the most salient 
features of the Bohm-Gadella theory with standard quantum mechanics.

Before turning to the main body of the paper, we would like to make three
remarks. First, although we will focus on the spherical shell potential for 
zero angular momentum, our results are valid for any partial wave and for a 
large class of potentials that include, in particular, potentials of finite 
range. The reason why our results are valid for such a large class of 
potentials
is that, ultimately, such results depend on whether one can analytically
continue the Jost and scattering functions into the whole complex plane. Since
such continuation is possible for potentials that fall off at infinity faster
than any exponential~\cite{NUSSENZVEIG}, our results remain valid for 
a whole lot of potentials. What is more, when the tails of the 
potential fall off slower than exponentials, the analytic properties of the
Jost and scattering functions are even in more disagreement with the 
Hardy axiom. Second, although the results of this paper render the rigged 
Hilbert space of Hardy class unsuitable for the Lippmann-Schwinger equation, 
the rigged Hilbert spaces of~\cite{LS1,LS2} lie at 
the heart of the foundations of the Lippmann-Schwinger equation. And third,
it is not that the math of the Bohm-Gadella theory is wrong, it is 
rather that the math of the Bohm-Gadella theory is inconsistent with quantum 
mechanics.

\section{Brief review of Hardy functions}
\label{sec:HFreview}

In this section, we list the definition and main properties of
functions of Hardy class.

\subsection{General properties of Hardy functions}
\label{sec:GenrpoHfun}

A Hardy function $f(z)$ on the upper half of the complex plane
${\mathbb C}^{+}$ is a function satisfying the following conditions
\cite{DUREN,HOFFMANN,KOOSISa,KOOSISb,TITCHMARSH}:
\begin{itemize}
   \item[({\it i})] It is analytic on the open upper
   half-plane, i.e., on the set of complex numbers with positive
   imaginary part.  
   \item[{\it (ii)}] For any value of $y>0$, the
   integral
\begin{equation}
      \int_{-\infty }^{+\infty} \rmd E \, |f(E+\rmi y)|^2 
        \label{68}
\end{equation}
converges.
\item[{\it (iii)}] For all $y>0$, these integrals are bounded by the
same constant $K$,
\begin{equation}
      \sup_{y>0}\int_{-\infty}^{+\infty}\rmd E \, |f(E+\rmi y)|^2 <K \, .
        \label{69}
\end{equation}
\end{itemize}

The set of Hardy functions on the upper half-plane, often referred to
as Hardy functions from above, is a linear space that we denote by
${\mathcal H}_{+}^2$.

Similarly, Hardy functions on the lower half-plane ${\mathbb C}^{-}$
are analytic on the open lower half-plane, and for these functions the
conditions~(\ref{68}) and (\ref{69}) hold with $y<0$. We denote the linear 
space of Hardy functions from below by ${\mathcal H}_{-}^2$.

Any Hardy function $f(z)$ has a unique boundary value $f(E)$ on the
real axis:
\begin{equation}
      \lim_{y\to 0}f(E \pm \rmi y)=f(E) \, , \quad f\in {\mathcal
      H}_{\pm}^2 \, .
\end{equation}  
The boundary value $f(E)$ is square integrable, and its squared norm is also 
bounded by $K$:
\begin{equation}
      \int_{-\infty}^{+\infty}\rmd E \, |f(E)|^2 <K \, .
        \label{69a}
\end{equation}
Thus, a function in ${\cal H}^2_\pm$ uniquely determines a square integrable 
function on $\mathbb R$.

An important theorem, due to Titchmarsh~\cite{TITCHMARSH}, states that
Hardy functions can be recovered from their boundary values on the real
line. If $f(E)$ is the function representing the boundary values of a
Hardy function $f(z)$ on ${\mathbb C}^{\pm }$, then
\begin{equation}
      f(z)=\pm \frac 1{2\pi \rmi}\int_{-\infty }^{+\infty} \rmd E
      \frac{f(E)}{E-z}  \, , 
           \label{70a}
\end{equation}
where the signs ($+$) and ($-$) correspond to Hardy functions on the
upper and lower half-planes, respectively. This one-to-one correspondence 
between the ${\cal H}_{\pm}^2$ functions and their boundary values on 
$\mathbb R$ allows the identification of $f(z)$ with $f(E)$ for 
$f\in {\cal H}_{\pm}^2$.

Another important theorem on Hardy functions is that by Paley and
Wiener~\cite{PALEY,DUREN,HOFFMANN,KOOSISa,KOOSISb}. The
theorem asserts that if $f \in {\cal H}_{+}^2$, then the Fourier transform
of $f$,
\begin{equation}
      \widetilde{f}(t) = 
             \frac{1}{\sqrt{2\pi}} \int_{-\infty}^{+\infty}\rmd E \, 
           \rme ^{-\rmi Et}f(E) \, ,  
      \label{FTPWthe}
\end{equation}
has the property $\widetilde{f}(t)=0$ for $t<0$. Similarly, if 
$f \in {\cal H}_{-}^2$, then the Fourier transform of $f$ has the property 
$\widetilde{f}(t)=0$ for $t>0$. The Paley-Wiener theorem is the major reason 
why Hardy functions have been widely used in causality problems in which the 
frequency (energy) runs through the entire real 
line~\cite{NUSSENZVEIG,VANKAMPEN1}. 

A theorem due to van Winter~\cite{VANWINTERa} establishes that a Hardy
function can be recovered from just its boundary values on the positive real
axis ${\mathbb R}^+$. Whether the recovered function is an element of
${\mathcal H}_{+}^2$ or ${\mathcal H}_{-}^2$ is determined by
means of the Mellin transform. Thus, if we call 
$\left. {\mathcal H}_{+}^2\right|_{{\mathbb R}^{+}}$
the space of boundary values on ${\mathbb R}^{+}$ of the functions in
${ \mathcal H}_{+}^2$, and 
$\left. {\mathcal H}_{-}^2\right|_{{\mathbb R}^{+}}$ the space of boundary
values on ${\mathbb R}^{+}$ of the functions in ${\mathcal H}_{-}^2$,
we have the following bijection:
\numparts
\begin{eqnarray}
      &&\Theta {\mathcal H}_{+}^2 =
        \left. {\mathcal H}_{+}^2\right|_{{\mathbb R}^{+}}  , \label{75}
      \\ &&\Theta {\mathcal H}_{-}^2 =
          \left. {\mathcal H}_{-}^2\right|_{{\mathbb R}^{+}}   ,
      \label{76}
\end{eqnarray}
\endnumparts
where the image of any $f_{\pm }(E)\in {\mathcal H}_{\pm }^2$ by
$\Theta$ is a function which is equal to $f_{\pm }(E)$ for 
$E \in {\mathbb R}^{+}$ and is not defined for negative values of $E$.

The following are among the other interesting properties of Hardy
functions~\cite{TITCHMARSH}:
\begin{itemize}
\item[{\it (i)}] Let us define the Hilbert transform for an 
$L^2({\mathbb R})$ function $f$ as
\begin{equation}
      \mathfrak{H}f(E)=\frac 1\pi \ {\mathcal P}\int_{-\infty }^{+\infty}
      \rmd x \,  \frac{f(x)}{x-E} \, , 
          \label{77}
\end{equation}
where ${\mathcal P}$ denotes the Cauchy principal value. The Hilbert
transform is linear and its image also lies in $L^2({\mathbb R})$. A
square integrable complex function $f(E)$, with real part $u(E)$ and
imaginary part $v(E)$, belongs to ${\mathcal H}_{\pm}^2$ if and only
if
\begin{equation}
        \mathfrak{H}u=\pm v\quad{\rm and}\quad\mathfrak{H}v={\mp}u \,
        .  \label{78}
\end{equation}
In particular, a Hardy function cannot be either real or purely
imaginary on the whole real line.

\item[{\it (ii)}] From ({\it i}), we immediately see that 
$f(E)\in {\mathcal H}_{\pm}^2$ if and only if its complex conjugate
$f^{*}(E)\in {\mathcal H}_{\mp}^2$.

\item[{\it (iii)}] A function $f$ of ${\mathcal H}_{+}^2$ (respectively
${\mathcal H}_{-}^2$) vanishes in the infinite arc of the upper (respectively
lower) half-plane:
\begin{equation}
       \lim_{z\to \infty}f(z) = 0 \, , \qquad  f \in {\cal H}_{\pm} \, , \
             z\in {\mathbb C}_{\pm} \, .
       \label{limiifn}
\end{equation}
More precisely, Hardy functions behave for large values of $z$ as 
$1/\sqrt z$ (cf.~\cite{KOOSISb}).

\item[{\it (iv)}] The spaces ${\mathcal H}_{+}^2$ and 
${\mathcal H}_{-}^2$ have a trivial intersection,
\begin{equation}
        {\mathcal H}_{+}^2 \cap {\mathcal H}_{-}^2 = \{ 0 \} \, .
\end{equation}
However, the spaces
of functions which are restrictions of Hardy functions to the positive
semiaxis ${\mathbb R}^{+}$, 
$\left. {\mathcal H}_{+}^2\right|_{{\mathbb R}^{+}}$ and 
$\left. {\mathcal H}_{-}^2\right|_{{\mathbb R}^{+}}$,
have a nontrivial intersection~\cite{BG},
\begin{equation}
       \left. {\mathcal H}_{+}^2\right|_{{\mathbb R}^{+}} \cap 
       \left. {\mathcal H}_{-}^2\right|_{{\mathbb R}^{+}}  \neq  \{ 0 \} \, .
\end{equation}
Moreover, the intersection 
$\left. {\mathcal H}_{+}^2\right|_{{\mathbb R}^{+}} \cap 
\left. {\mathcal H}_{-}^2\right|_{{\mathbb R}^{+}}$ is dense in 
$L^2({\mathbb R}^{+})$.
\end{itemize}

\subsection{The rigged Hilbert spaces of Hardy class}

Two pairs of rigged Hilbert spaces of Hardy class form the backbone of the
Bohm-Gadella theory. The first pair is given by~\cite{BG}
\begin{equation}
      {\mathcal S}\cap {\mathcal H}_{\pm }^2\subset {\mathcal H}_{\pm}^2
       \subset ({\mathcal S}\cap {\mathcal H}_{\pm }^2)^{\times }   , 
      \label{79}
\end{equation}
where $\cal S$ denotes the Schwartz space on the real line. The second
pair is constructed as follows. We mentioned earlier that Hardy functions are
determined by their values on the positive real axis plus a
specification which says if they are Hardy on the upper or the lower
half-planes. Thus, we have defined the spaces 
$\Theta{\mathcal H}_{+}^2=\left. {\mathcal H}_{+}^2\right|_{{\mathbb R}^{+}}$
and 
$\Theta{\mathcal H}_{-}^2=\left. {\mathcal H}_{-}^2\right|_{{\mathbb R}^{+}}$.
Now consider
\numparts
\begin{eqnarray}
     &&\left. {\mathcal S}\cap {\mathcal H}_{+}^2\right| _{{\mathbb
     R}^{+}}= \Theta ({\mathcal S}\cap {\mathcal H}_{+}^2) \, ,
     \label{80} \\ 
   &&\left. {\mathcal S}\cap {\mathcal H}_{-}^2\right|
     _{{\mathbb R}^{+}} =\Theta ({\mathcal S} \cap {\mathcal
     H}_{-}^2) \, . 
        \label{81}
\end{eqnarray}
\endnumparts
The spaces $\left. {\mathcal S}\cap {\mathcal H}_{\pm }^2\right|
_{{\mathbb R}^{+}}$ are dense in $L^2({\mathbb R}^{+})$ and
yield the second pair of rigged Hilbert spaces of Hardy class~\cite{BG}:
\begin{equation}
    \left. {\mathcal S}\cap {\mathcal H}_{\pm }^2\right| _{{\mathbb
    R}^{+}} \subset L^2({\mathbb R}^{+})\subset (\left. {\mathcal
    S}\cap {\mathcal H}_{\pm }^2\right| _{{\mathbb R}^+} )^{\times}  .
    \label{82}
\end{equation}

\subsection{The Hardy axiom}

If we denote the Lippmann-Schwinger bras by $\langle ^{\pm}E|$, and if we 
denote the ``in'' and the ``out'' wave functions by $\varphi ^{+}$ and 
$\psi ^-$, then the Hardy axiom states that   
\begin{equation}
      \langle ^{+}E|\varphi ^{+}\rangle = \varphi ^{+}(E) \in  
      \left. {\mathcal S}\cap {\mathcal H}_{-}^2\right| _{{\mathbb R}^{+}} ,
\end{equation}
\begin{equation}
      \langle ^{-}E|\psi ^{-}\rangle = \psi ^{-}(E) \in  
      \left. {\mathcal S}\cap {\mathcal H}_{+}^2\right| _{{\mathbb R}^{+}} .
\end{equation}
By~(\ref{limiifn}), these assumptions imply, in particular, that the analytic 
continuation
of $\varphi ^{+}(E)$ and $\psi ^{-}(E)$ tend to zero in, respectively,
the lower and upper infinite arcs of the second sheet of the Riemann surface:
\begin{equation}
       \lim_{z\to \infty} \varphi ^{+}(z) = 0 \, , \quad
     z\in {\mathbb C}_{\rm II}^{-}  \, , \qquad 
            \mbox{if $\varphi ^{+}(E)$ was a Hardy function,}
         \label{asymlim+}
\end{equation}
\begin{equation}
       \lim_{z\to \infty} \psi ^{-}(z) = 0 \, , \quad
     z\in {\mathbb C}_{\rm II}^{+}     \, , \qquad
          \mbox{if $\psi ^{-}(E)$ was a Hardy function,}
          \label{asymlim-}
\end{equation}
where ${\mathbb C}_{\rm II}^{\pm}$ denote the upper ($+$) and the lower ($-$)
half-planes of the second sheet. As we will see in the next section,
in quantum mechanics these limits are not zero, and therefore 
$\varphi ^{+}(E)$ and $\psi ^{-}(E)$ are not Hardy functions.

\section{The asymptotics of the Lippmann-Schwinger equation vs.~the asymptotics
of the Bohm-Gadella theory}
\label{sec:alinfar}

We are going to see that the limits~(\ref{asymlim+}) and (\ref{asymlim-}) are 
in general not zero by way of a simple counterexample. We will focus
on the limit~(\ref{asymlim+}), since the same arguments apply to the
limit~(\ref{asymlim-}).

The simple counterexample we will use is the spherical shell potential:
\begin{equation}
        V({\bf x})\equiv V(r)=\left\{ \begin{array}{ll}
                                0   &0<r<a  \\
                                V_0 &a<r<b  \\
                                0   &b<r<\infty \, ,
                  \end{array} 
                 \right. 
	\label{sbpotential}
\end{equation}
where $V_0$ is a positive constant that determines the strength of the
potential, and $a$ and $b$ determine the positions in between which the
potential is nonzero.

We will work in the radial, position representation for zero angular 
momentum and take $\hbar ^2/(2m)=1$. In this representation, the Hamiltonian 
$H$ acts as
\begin{equation}
      H = -\frac{\rmd ^2}{\rmd r^2} + V(r)    \, .  
  \label{doh}
\end{equation}
The Lippmann-Schwinger eigenfunctions are given by~\cite{LS1,LS2,TAYLOR,NEWTON}
\begin{equation}
      \langle r|E^{\pm}\rangle \equiv 
       \chi ^{\pm}(r;E)= N(E) \,  
     \frac{\chi (r;E)}{{\cal J}_{\pm}(E)}  \, ,
      \label{pmeigndu}
\end{equation}
where $N(E)$ is a delta-normalization factor,
\begin{equation}
      N(E)=\sqrt{\frac{1}{\pi}
         \frac{1}{\sqrt{E \,}\,}\,} \, ,
        \label{Nfactor}
\end{equation}
$\chi (r;E)$ is the so-called regular solution,
\begin{equation}
     \hskip-1cm  \chi (r;E) = \left\{
             \begin{array}{ll}
                  \sin ( \sqrt{E\,}\,r)  & 0<r<a \\
               {\cal J}_1(E)\rme ^{\rmi \sqrt{E-V_0 \,}\,r}
              +{\cal J}_2(E)\rme^{-\rmi\sqrt{E-V_0 \,}\,r}
                                                         & a<r<b \\ 
                  {\cal J}_3(E) \rme^{\rmi \sqrt{E \,}\,r}
                   +{\cal J}_4(E) \rme^{-\rmi\sqrt{E\,}\,r}
                                                         & b<r<\infty \, ,
               \end{array}
              \right.
        \label{LSchi}
\end{equation}
and ${\cal J}_{\pm}(E)$ are the Jost functions,
\numparts
\begin{equation}
        {\cal J}_+(E)=-2\rmi {\cal J}_4(E) \, ,
\end{equation}
\begin{equation}
        {\cal J}_-(E)=2\rmi {\cal J}_3(E) \, .
\end{equation}
\endnumparts
The explicit expressions for ${\cal J}_1$-${\cal J}_4$ can be obtained by 
matching the values of $\chi (r;E)$ and of its derivative at the 
discontinuities of the potential.

If we denote the ``in'' wave function in the position representation by 
$\varphi ^+(r)$, then the ``in'' wave function in the energy representation, 
$\varphi ^+(E)$, is given by~\cite{LS1}
\begin{equation}
       \varphi ^+(E) = 
           \int_0^{\infty}\rmd r \, \varphi ^+(r) \overline{\chi ^+(r;E)} =
           \int_0^{\infty}\rmd r \,  \varphi ^+(r) \chi ^-(r;E) \, .
      \label{LSinteexpre+}
\end{equation}
The analytic continuation of $\varphi ^+(E)$ into the complex plane can
be readily obtained by analytically continuing 
$\chi ^-(r;E)$~\cite{LS2}. The resulting analytic continuations are
denoted by $\varphi ^+(z)$ and $\chi ^-(r;z)$:
\begin{equation}
       \varphi ^+(z) = 
           \int_0^{\infty}\rmd r \,  \varphi ^+(r) \chi ^-(r;z) \, .
      \label{LSinteexpre+a}
\end{equation}

In order to show that the limit~(\ref{asymlim+}) does not hold, we will
compare that limit with the pace at which $\varphi ^+(z)$ grows in the 
lower half-plane of the second sheet. From Eq.~(\ref{LSinteexpre+a}) and
from the results of Ref.~\cite{LS2}, we know that such a pace is 
determined by the falloff of $\varphi ^+(r)$ as $r$ tends to 
infinity and by the growth of $\chi ^-(r;z)$, which we now obtain.

The growth of $\chi ^-(r;z)$ is determined by the growths of the regular 
solution and of the inverse of ${\cal J}_-(z)$, as can be seen in 
Eq.~(\ref{pmeigndu}). In its turn, the growth of the regular solution is 
given by the following estimate (see, for example, Eq.~(12.6) in 
Ref.~\cite{TAYLOR}):
\begin{equation}
      \left| \chi (r; z)\right| \leq C \, 
        \frac{\left|z\right|^{1/2}r}{1+\left|z\right|^{1/2}r} \,  
      \rme ^{|{\rm Im}\sqrt{z\,}|r} \, , \quad  
                        z\in {\mathbb C} \, ,   
      \label{boundrs}
\end{equation}
where $C$ is a constant. In the lower half-plane of the second sheet, the 
growth of $1/{\cal J}_-(z)$ can be shown to be bounded by a 
constant~\cite{LS2}:
\begin{equation}
      \left| \frac{1}{{\cal J}_-(z)} \right| \leq C  \, , \quad  
                        z \in {\mathbb C}_{\rm II}^-   \, .   
      \label{boundJf}
\end{equation}
The combination of the last two inequalities with Eq.~(\ref{pmeigndu}) yields
the growth of $\chi ^-(r;z)$ in the lower half-plane of the second sheet:
\begin{equation}
      \left| \chi ^-(r;z)\right| \leq C \, 
        \frac{\left|z\right| ^{1/4} r}{1+\left|z\right|^{1/2}r} \,  
      \rme ^{|{\rm Im}\sqrt{z\,}|r}  , \quad  
          z  \in {\mathbb C}_{\rm II}^-   \, . 
      \label{boundLS-}
\end{equation}
Hence, when $z \in {\mathbb C}_{\rm II}^-$, $\chi ^-(r;z)$ blows up 
{\it exponentially} as $z$ tends to infinity. 

By using the estimate~(\ref{boundLS-}) and the Gelfand-Shilov theory
of $M$ and $\Omega$ functions~\cite{GELFANDIII}, one can obtain
the growth of $\varphi ^+(z)$ from the falloff of $\varphi ^+(r)$, 
see~\cite{LS2}. The result is that if $a$ and $b$ are positive real numbers
satisfying
\begin{equation}
       \frac{1}{a}+\frac{1}{b} = 1 \, ,
\end{equation}
and if $\varphi ^+(r)$ is an infinitely differentiable function
whose tails fall off like $\rme ^{-r^a/a}$, then $\varphi ^+(z)$ grows
like $\rme ^{|{\rm Im}(\sqrt{z})|^b/b}$ in the infinite arc of the energy 
plane:
\begin{equation}
       {\rm If} \ \varphi ^+(r) \sim \rme ^{-\frac{\, r^a}{a}}
        \  {\rm as} \ r \to \infty , \ {\rm then} \
      \varphi ^+(z) \sim \rme ^{\frac{\, |{\rm Im}(\sqrt{z})|^b}{b}} \
       {\rm as} \ z\to \infty  \,.
        \label{blowupfex}
\end{equation}
Thus, when in the position representation a function has tails, in the
energy representation it blows up faster than any exponential, and therefore
it cannot be a Hardy function, since the blowup~(\ref{blowupfex}) contradicts
the limit~(\ref{asymlim+}) demanded by the Hardy axiom.

One could think of saving the limit~(\ref{asymlim+}) by imposing further 
restrictions on the falloff of $\varphi ^+(r)$ at infinity, because such
restrictions slow down the growth of $\varphi ^+(z)$. The strongest 
restriction of
this kind is to impose that $\varphi ^+(r)$ be an infinitely differentiable
function with compact support, $\varphi ^+(r) \in C_0^{\infty}$. However, 
as we are going to see, functions of compact support do not 
comply with the limit~(\ref{asymlim+}) either.

Let us take $\varphi ^+(r) \in C_0^{\infty}$ such that
$\varphi ^+(r)=0$ when $r>A$. Then,
\begin{eqnarray}
     \left| \varphi ^+(z) \right| &= 
        \left| \int_0^{\infty}\rmd r \,  \varphi ^+(r) \chi ^-(r;z) \right|   
             & \qquad  \mbox{by (\ref{LSinteexpre+a})}  \nonumber \\
    & \leq \int_0^{\infty}\rmd r \,  \left| \varphi ^+(r) \right| 
                  \left| \chi ^-(r;z) \right|  & \quad \nonumber \\
    & =  \int_0^{A}\rmd r \,  \left| \varphi ^+(r) \right| 
                  \left| \chi ^-(r;z) \right|  & \quad \nonumber \\
    & \leq C \, 
        \frac{\left|z\right| ^{1/4} A}{1+\left|z\right|^{1/2}A} \,  
      \rme ^{|{\rm Im}\sqrt{z\,}|A}  
          \int_0^{A}\rmd r \,  \left| \varphi ^+(r) \right| 
                   & \qquad  \mbox{by (\ref{boundLS-})}  \nonumber \\
     & \leq C \, 
        \frac{\left|z\right| ^{1/4} A}{1+\left|z\right|^{1/2}A} \,  
      \rme ^{|{\rm Im}\sqrt{z\,}|A}  \, ; 
                   & \quad 
      \label{stepsforlous}
\end{eqnarray}
that is, when $\varphi ^+(r) \in C_0^{\infty}$, $\varphi ^+(z)$ blows up 
{\it exponentially} in the infinite arc of ${\mathbb C}_{\rm II}^-$:
\begin{equation} 
     {\rm If} \  |\varphi ^+(r)| = 0 \ {\rm when} \ r>A , \ 
      {\rm then} \ \varphi ^+(z) \sim \rme ^{A|{\rm Im}\sqrt{z\,}|} 
       \ {\rm as} \ z\to \infty  \, .
        \label{grofainf}
\end{equation}
Quite a contrast to the limit~(\ref{asymlim+}). 

Thus, no matter how small the tails of the wave functions in the position 
representation are, in the energy representation the wave functions blow 
up at least {\it exponentially} in the infinite arc of 
${\mathbb C}_{\rm II}^-$, in contradiction to the limit~(\ref{asymlim+})---the 
Hardy axiom is inconsistent with the Lippmann-Schwinger equation.

\section{The preparation-registration arrow of time}
\label{sec:PRAT}

The preparation-registration arrow of time provides the physical 
justification for the Hardy axiom. In this section, we 
will demonstrate that the derivation of the preparation-registration arrow 
of time is flawed, thereby demonstrating that the Hardy axiom lacks a physical
basis.

The derivation of the preparation-registration arrow of time can be found,
for example, in section~3 of Ref.~\cite{PLA94} and in section~III of
Ref.~\cite{JMP95}, and it goes as follows. A scattering experiment consists 
of a preparation stage and a registration stage. In the preparation stage, a
beam is prepared in an initial state
$\varphi ^{\rm in}$ before the interaction $V=H-H_0$ is turned
on. The initial state $\varphi ^{\rm in}$ evolves according to the free
Hamiltonian $H_0$, 
$\varphi ^{\rm in}(t) =\rme ^{-\rmi H_0t}\varphi ^{\rm in}$. When the beam
reaches the interaction region, the free initial state $\varphi ^{\rm in}$
turns into the exact ``in'' state $\varphi ^{+}$, which evolves 
according to the total Hamiltonian $H$, 
$\varphi ^{+}(t) = \rme ^{-\rmi Ht}\varphi ^{+}$. The beam then leaves the 
interaction region and ends up as the state $\varphi ^{\rm out}$.

In the registration stage, the detector outside the interaction region does
in general not detect $\varphi ^{\rm out}$ but rather the overlap of 
$\varphi ^{\rm out}$ with a final state $\psi ^{\rm out}$. The state 
$\psi ^{\rm out}$ evolves according to the free Hamiltonian and corresponds
to an exact ``out'' state $\psi ^-$. The ``out'' state $\psi ^-$ evolves 
according to the total Hamiltonian. 

A well-known result of scattering theory is that the states
$\varphi ^{\rm in}$, $\varphi ^{+}$ and $\psi ^{\rm out}$, $\psi ^{-}$
satisfy 
\begin{equation}
     \varphi ^{\rm in}(E)= \langle E|\varphi ^{\rm in}\rangle =  
        \langle ^+E|\varphi ^{+}\rangle = \varphi ^{+}(E)  \, ,
\end{equation}
\begin{equation}
    \psi ^{\rm out}(E) =  \langle E|\psi ^{\rm out}\rangle =  
     \langle ^-E|\psi ^{-}\rangle = \psi ^- (E) \, ,
\end{equation}
where $\langle E|$ are the eigenbras of $H_0$, and $\langle ^{\pm}E|$ are
the Lippmann-Schwinger bras.

One now takes $t=0$ as the time before which the preparation of 
$\varphi ^{\rm in}$ is completed, and after which the registration of
$\psi ^{\rm out}$ begins. Then, as the mathematical statement for ``no 
preparations for $t>0$,'' one takes
\begin{equation}
     |\langle E|\varphi ^{\rm in}(t)\rangle|=  
      |\langle ^+E|\varphi ^{+}(t)\rangle|=  0 \, , \quad t>0 \, , 
        \label{flawassum} 
\end{equation}
for all energies, which implies
\begin{equation}
    \hskip-1.5cm
    0= \int_{-\infty}^{+\infty}\rmd E \, 
          \langle E|\varphi ^{\rm in} (t)\rangle =  
     \int_{-\infty}^{+\infty}\rmd E \,      
              \langle ^+E|\varphi ^{+}(t) \rangle =  
     \int_{-\infty}^{+\infty}\rmd E \,      
              \langle ^+E|\rme ^{-\rmi Ht} |\varphi ^{+} \rangle  \, ,
\end{equation}
or
\begin{equation}
    0 = \int_{-\infty}^{+\infty}\rmd E \,   
       \rme ^{-\rmi Et} \varphi ^{+}(E) \equiv 
       \widetilde{\varphi}^+(t) \, , \quad \mbox{for} \ t>0  \, .
      \label{ferFtvph+}
\end{equation}
As the mathematical statement for ``no registrations for $t<0$,'' one takes 
\begin{equation}
     |\langle E|\psi ^{\rm out}(t)\rangle|=
      |\langle ^-E|\psi ^{-}(t)\rangle|=  0 \, , \quad t<0 \, , 
\end{equation}
for all energies, which implies
\begin{equation}
     \hskip-1.7cm
    0= \int_{-\infty}^{+\infty}\rmd E \, 
          \langle E|\psi ^{\rm out} (t)\rangle =  
       \int_{-\infty}^{+\infty}\rmd E \,
        \langle ^-E|\psi ^{-}(t) \rangle =
     \int_{-\infty}^{+\infty}\rmd E \,
        \langle ^-E|\rme^{-\rmi Ht}|\psi ^{-} \rangle  \, ,
\end{equation}
or
\begin{equation}
    0 = \int_{-\infty}^{+\infty}\rmd E \,   
       \rme ^{-\rmi Et} \psi ^{-}(E) \equiv 
       \widetilde{\psi}^-(t)  \, , \quad \mbox{for} \ t<0   \, .
    \label{ferFtvps-}
\end{equation}
By the Paley-Wiener theorem, Eq.~(\ref{ferFtvph+}) implies that
$\varphi ^+(E)=\varphi ^{\rm in}(E)$ is a Hardy function from below, and
Eq.~(\ref{ferFtvps-}) implies that $\psi ^-(E)=\psi ^{\rm out}(E)$ is a Hardy 
function from above. Hence the claim that the Hardy axiom is
rooted in causality principles.

We are now in a position to reveal the flaws of the preparation-registration
arrow of time. For the sake of simplicity, we will focus on the ``in'' 
states, since the same arguments apply to the ``out'' states.

The first flaw lies in assumption~(\ref{flawassum}). From such an assumption,
it follows that
\begin{equation}
     0=\langle ^+E|\varphi ^{+}(t)\rangle = \rme ^{-\rmi Et} \varphi ^{+}(E) 
        \label{flawassum1} 
\end{equation}
for all energies. Hence, 
\begin{equation}
     0=\varphi ^{+}(E)
        \label{flawassum2} 
\end{equation}
for all energies, which can happen only when $\varphi ^+$ is identically 
0. Thus, the preparation-registration arrow of time holds only in the
meaningless case of the zero wave function.

Condition~(\ref{ferFtvph+}), on which condition the Hardy axiom rests,
is weaker than the flawed assumption~(\ref{flawassum}). One may then think
of saving the Hardy axiom by assuming simply Eq.~(\ref{ferFtvph+}), which 
by the Paley-Wiener theorem is equivalent to assuming that $\varphi ^+(E)$ 
is a Hardy function from below. This way of stating the Hardy axiom,
which is the way used in Refs.~\cite{FP03a,IJTP03}, is tantamount
to the causal condition ``$\widetilde{\varphi}^+(t)=0$ for $t>0$.'' The
problem is that this causal condition is dead-end, because the 
quantum mechanical time evolution of $\varphi ^+$,
\begin{equation}
      \rme ^{-\rmi Ht}\varphi ^+ = \int_0^{\infty}\rmd E \,   
         \rme ^{-\rmi Et}  |E^+\rangle \langle ^+E|\varphi ^+\rangle \, ,
      \label{QMtimevo}
\end{equation}
is {\it different} from $\widetilde{\varphi}^+(t)$:
\begin{equation}
      \varphi ^+(t)= \rme ^{-\rmi Ht} \varphi ^+ \ \neq \   
       \widetilde{\varphi}^+(t)  \, .
\end{equation}
It is clear from this equation that any statement on the the causal behavior 
of $\widetilde{\varphi}^+(t)$ will in general not apply to the true quantum 
mechanical evolved state $\varphi ^+(t)$ (even if $\varphi ^+$ were a Hardy 
function). Therefore, there is no reason why the solutions of the 
Lippmann-Schwinger equation should be related to Hardy functions.

One could have foreseen that the causality principles of the Hardy axiom 
are inconsistent with quantum mechanics by recalling that in standard 
scattering theory, a beam is prepared at $t=-\infty$, hits the
target, and outgoes to infinity at $t=+\infty$, and therefore its time 
evolution lasts from $t=-\infty$ till $t=+\infty$.

Three remarks are in order here. (1) Because $\rme ^{-\rmi Ht}$ is unitary,
$\rme ^{-\rmi Ht} \varphi ^+=0$ if and only if $\varphi ^+ \equiv 0$. Thus, 
there is little hope of having $\varphi ^+(t)=0$ for any $t>0$, since such
condition forces $\varphi ^+$ to vanish. (2) If the argument behind
the preparation-registration arrow of time was physically sound, it would 
apply to any quantum mechanical energy distribution, not just to those
associated with the solutions
of the Lippmann-Schwinger equation. Also, it would apply to any system, be 
it quantum or classical. 
(3) The entire real line of 
energies has been used in Eq.~(\ref{ferFtvph+}), whereas the 
quantum mechanical time evolution~(\ref{QMtimevo}) uses the physical 
spectrum $[0, \infty )$.

\section{The spectrum of the Hamiltonian is bounded from below}
\label{sec:SpectrumH}

That the Paley-Wiener theorem makes use of the entire real line of 
energies through the Fourier transform~(\ref{FTPWthe}) seems to indicate 
that Hardy functions fit systems where energy (frequency) integrations
are taken over the whole real line, 
rather than quantum mechanical systems, the majority of which have spectra 
bounded from below. This important point is known to some 
authors. For example, Nussenzveig~\cite{NUSSENZVEIG} and van 
Kampen~\cite{VANKAMPEN1} utilize ${\cal H}_{\pm}^2$ functions in causality 
problems of classical electromagnetic scattering, where the frequency 
integrals are taken over the entire real line. However, both 
Nussenzveig~\cite{NUSSENZVEIG} and van Kampen~\cite{VANKAMPEN2} avoid the use 
of ${\cal H}_{\pm}^2$ functions in
non-relativistic scattering, where the spectrum is bounded from below. In 
this section, we are going to delve into why Hardy functions are not
suitable for systems whose spectrum is bounded from below. 

We start by recalling a basic result of the spectral theory of linear 
operators on a Hilbert space. Let $A$ denote a self-adjoint operator on a 
Hilbert space $\cal H$, and let ${{\rm Sp}(A)}$ denote its spectrum. For 
simplicity, let us assume that the spectrum of $A$ is purely 
continuous. Associated with each $\lambda \in {{\rm Sp}(A)}$, there is
a ket $|\lambda \rangle$ that is a (generalized) eigenvector of $A$:
\begin{equation}
    A|\lambda \rangle = \lambda |\lambda \rangle \, .
  \label{Aeieque}
\end{equation}
In practical applications, Eq.~(\ref{Aeieque}) is solved in the position 
representation:
\begin{equation}
    \langle x|A|\lambda \rangle = \lambda  \langle x|\lambda \rangle \, .
  \label{Aeiequexre}
\end{equation}
This equation has in general solutions for any complex $\lambda$, even
though the spectrum of a self-adjoint operator is real. The $\lambda$'s of 
${{\rm Sp}(A)}$ are distinguished from the rest of real and complex numbers
by the fact that their corresponding 
eigenfunctions $\langle x|\lambda \rangle$ are polynomially 
bounded~\cite{GALINDO}. The solutions of Eq.~(\ref{Aeiequexre}) for 
$\lambda \in  {{\rm Sp}(A)}$ can then be used to express the norm of a 
sufficiently smooth function $f$ and the expectation of $A$ in $f$:
\begin{equation}
    (f,f) = \int_{{\rm Sp}(A)} \rmd \lambda \,  
                \langle f|\lambda \rangle \langle \lambda | f\rangle  =
   \int_{{\rm Sp}(A)} \rmd \lambda \,
                |f(\lambda)|^2 < \infty  \, ,
          \label{spfg}
\end{equation}
\begin{equation}
    (f,Af) = \int_{{\rm Sp}(A)} \rmd \lambda \, \lambda 
                \langle f|\lambda \rangle \langle \lambda | f\rangle 
           = \int_{{\rm Sp}(A)} \rmd \lambda \, \lambda 
                |f(\lambda ) |^2 < \infty  \, .
          \label{speA}
\end{equation}
It is important to note that in these integrals, $\lambda$ runs over the 
physical spectrum.

The eigenfunctions of the Hamiltonian~(\ref{doh}) are given by 
Eq.~(\ref{pmeigndu}). As Eq.~(\ref{boundrs}) shows, the 
eigenfunctions~(\ref{pmeigndu}) are polynomially bounded only 
when $E\geq 0$, and therefore the spectrum of $H$ is the positive real 
line. Thus, when we apply the last two equations to $H$ and $\varphi ^+$, 
we obtain
\begin{equation}
    \hskip-1cm (\varphi ^+,\varphi ^+) = \int_0^{\infty} \rmd E \, 
                \langle \varphi ^+|E^+ \rangle \langle ^+E|\varphi ^+\rangle 
           = \int_0^{\infty} \rmd E \, |\varphi ^+(E)|^2 < \infty \, ,
      \label{scpphs}
\end{equation}
\begin{equation}
   \hskip-1cm  (\varphi ^+,H\varphi ^+) = \int_0^{\infty} \rmd E \, 
               E \langle \varphi ^+|E^+ \rangle \langle ^+E|\varphi ^+\rangle 
           = \int_0^{\infty} \rmd E \, E |\varphi ^+(E)|^2 < \infty \, .
     \label{expvalh}
\end{equation}
Note that in these two integrals, the energy ranges from $0$ to $\infty$.

We now recall that the Hardy axiom implies that the following
integrals over the entire real line are finite:
\begin{equation}
    \hskip-2cm (\varphi ^{+},\varphi ^{+})_{\rm Hardy} = 
          \int_{-\infty}^{+\infty} \rmd E \,  
     \langle \varphi ^{+}|E ^+\rangle \langle ^+E|\varphi ^{+}\rangle 
           = \int_{-\infty}^{+\infty} \rmd E \,  
                |\varphi ^{+}(E) |^2 < K <\infty  \, ,
          \label{Haxconneg1}
\end{equation}
\begin{equation}
    \hskip-2cm (\varphi ^{+},H\varphi ^{+})_{\rm Hardy} = 
          \int_{-\infty}^{+\infty} \rmd E \, E 
     \langle \varphi ^{+}|E ^+\rangle \langle ^+E|\varphi ^{+}\rangle 
           = \int_{-\infty}^{+\infty} \rmd E \, E 
                |\varphi ^{+}(E) |^2  <\infty  \, .
          \label{Haxconneg2}
\end{equation}
From Eqs.~(\ref{scpphs})-(\ref{Haxconneg2}) it follows that
\begin{equation}
     (\varphi ^{+},\varphi ^{+})_{\rm Hardy} = 
      (\varphi ^{+},\varphi ^{+}) +
        \int_{-\infty}^{0} \rmd E \,  
     \langle \varphi ^{+}|E ^+\rangle \langle ^+E|\varphi ^{+}\rangle \, ,
          \label{Haxconneg1a}
\end{equation}
\begin{equation}
      (\varphi ^{+},H\varphi ^{+})_{\rm Hardy} = 
          (\varphi ^{+},H\varphi ^{+}) +  
          \int_{-\infty}^{0} \rmd E \, E 
     \langle \varphi ^{+}|E ^+\rangle \langle ^+E|\varphi ^{+}\rangle  \, .
          \label{Haxconneg2a}
\end{equation}
Thus, if the Hardy axiom held, the integrals over the negative
real line would be finite. As we are going to see, this is not so.

When $\varphi ^{+}(r) \in C_0^{\infty}$ and $E\in (-\infty ,0)$, the 
estimate~(\ref{stepsforlous}) becomes
\begin{equation}
     \left| \varphi ^+(E) \right|  \leq C \, 
        \frac{\left|E\right| ^{1/4} A}{1+\left|E\right|^{1/2}A} \,  
      \rme ^{|E|^{1/2}A} \, , \quad E < 0  \, ,
      \label{stepsforlousenga}
\end{equation}
and therefore $\varphi ^+(E)$ blows up exponentially when $E$ tends to 
$-\infty$,
\begin{equation}
     \varphi ^+(E)  \sim \rme ^{|E|^{1/2}A} \, , 
           \quad E \to - \infty  \, .
      \label{stepsforlouseng}
\end{equation}
In addition, as shown in Sec.~\ref{sec:alinfar}, when $\varphi ^+(r)$ has 
non-vanishing tails, the growth of $\varphi ^+(z)$ 
outpaces~(\ref{stepsforlouseng}). Thus, the following integrals do not 
converge for very many reasonable wave functions: 
\begin{equation}
         \int_{-\infty}^0 \rmd E \, 
                \langle \varphi ^+|E^+\rangle \langle ^+E | \varphi ^+ \rangle 
           = \int_{-\infty}^0 \rmd E \, 
                |\varphi ^+ (E) |^2 = \infty   \, ,
     \label{dineint1}
\end{equation}
\begin{equation}
        \int_{-\infty}^0 \rmd E \, E 
                \langle \varphi ^+|E^+\rangle \langle ^+E | \varphi ^+ \rangle 
           = \int_{-\infty}^0 \rmd E \, E 
                |\varphi ^+ (E) |^2 = \infty   \, ,
     \label{dineint2}
\end{equation}
and
\begin{equation}
    (\varphi ^{+},\varphi ^{+})_{\rm Hardy} = \infty \, , 
          \label{Haxconneg1b}
\end{equation}
\begin{equation}
    (\varphi ^{+},H\varphi ^{+})_{\rm Hardy}  = \infty  \, ,
          \label{Haxconneg2b}
\end{equation}
in contradiction with the Bohm-Gadella assumptions~(\ref{Haxconneg1}) and 
(\ref{Haxconneg2}).

Note that if the spectrum of $H$ were the entire real line, the
Bohm-Gadella assumptions~(\ref{Haxconneg1})-(\ref{Haxconneg2}) would 
hold, because in such case the integrals in Eqs.~(\ref{spfg})-(\ref{speA})
would be taken over the full real line. Hence, the Hardy axiom might be
suitable for systems whose spectrum is the entire real line, although it is 
definitely not suitable for systems whose spectrum is bounded from 
below. Nevertheless, even if the Hardy axiom were allowed for Hamiltonians 
whose spectrum is the whole real line, such Hamiltonians are very rear, and 
they cannot be used in scattering theory, because the (continuous) spectrum
of a scattering Hamiltonian coincides with the spectrum of the kinetic 
energy operator, which is $[0,\infty )$.

The advocates of the Bohm-Gadella theory may argue that in their theory, they
integrate over the physical spectrum, negative-energy integrals
such as those in~(\ref{dineint1}) and (\ref{dineint2}) appear only by 
analytic continuation, and such negative-energy integrals are actually 
taken over the negative energy real line of the second sheet. The problem is 
that, although there is nothing wrong with analytically continuing from the 
physical spectrum into the negative real line of the second sheet, the 
resulting integrals are not convergent, as shown above. Assuming that 
they are convergent is in contradiction with the semiboundedness 
of the Hamiltonian.

\section{The Bohm-Gadella theory vs.~standard quantum mechanics}
\label{sec:GSdBSss}

The Bohm-Gadella theory deviates from standard quantum 
mechanics in many essential aspects. In this section, we point out the most 
important ones.

In the Bohm-Gadella theory, the time evolution of the Lippmann-Schwinger bras 
$\langle ^{\pm}E|$ and kets $|E^{\pm}\rangle$ is given by 
semigroups~\cite{BKW}, whereas in standard quantum 
mechanics it is given by a group~\cite{LS1}. In standard quantum mechanics, 
only the time evolution of the analytic continuation of
$\langle ^{\pm}E|$ and $|E^{\pm}\rangle$ is given by a semigroup~\cite{LS2}.

In the Bohm-Gadella theory, $\langle ^+E|$ and $|E^-\rangle$ are analytically
continued from the upper rim of the cut into the lower half-plane of the 
second sheet, and $\langle ^-E|$ and $|E^+\rangle$ are continued
from the lower rim of the cut into the upper half-plane of the second 
sheet~\cite{IJTP03}. By contrast, in standard quantum mechanics, 
$\langle ^{\pm}E|$ and $|E^{\pm}\rangle$ are all continued from the upper rim 
of the cut into the whole complex plane~\cite{LS2}.

In the Bohm-Gadella theory, the Lippmann-Schwinger bras and kets
act on two different rigged Hilbert spaces~\cite{BG}. In standard quantum 
mechanics, the Lippmann-Schwinger bras and kets act, in the position
representation, on one and the same rigged Hilbert space~\cite{LS1}.

In contrast to scattering theory, which seeks asymptotic completeness,
the Bohm-Gadella theory forgoes asymptotic completeness: 
\begin{quote}
``[$\cdots$] {\it we change just one axiom (Hilbert space 
and/or asymptotic completeness) to a new axiom which distinguishes between 
(in-)states and (out)observables using Hardy spaces.}'' (See the abstract 
and page~2324 of~\cite{IJTP03}.) 
\end{quote}
The Bohm-Gadella theory still uses the M{\o}ller operators and the $S$ 
matrix, although it is not explained how it is possible to have them without
asymptotic completeness.

In the Bohm-Gadella theory,
\begin{quote}
``[$\cdots$] {\it the Hamiltonian operator and boundary 
conditions alone do not specify a resonance state.}'' 
(See~\cite{IJTP03}, page~2332.)
\end{quote}
By contrast, in standard resonance theory, the Gamow states are obtained by 
solving the Schr\"odinger equation subject to purely outgoing boundary 
conditions~\cite{02AJP}.

In the Bohm-Gadella theory, the Lippmann-Schwinger equation is never solved,
and the properties of its solutions are assumed and promoted to an
axiom. This (Hardy) axiom becomes the cornerstone of the theory:
\begin{quote}
``{\it The unified theory of resonances and decay 
requires this new axiom.}'' (See~\cite{IJTP03}, page~2333.)
\end{quote}
By contrast, as shown in the present paper, the Hardy axiom is not only 
unnecessary but also inconsistent with the standard quantum mechanical 
theory of resonances and decay.

On top of the above problems, the Bohm-Gadella theory lacks a simple quantum 
mechanical scattering system to which it applies.

\section{The ``Gamow vectors'' and time asymmetry}
\label{sec:GSdBS}

The Bohm-Gadella theory was devised to accommodate resonance states in a
well-defined manner. Time asymmetry was a byproduct. Since resonance states 
and time asymmetry are two important aspects of quantum mechanics, and since 
the Bohm-Gadella theory 
has turned out to be flawed, one may wonder if those two aspects can be 
described within ordinary quantum mechanics. Fortunately, they can.

Time asymmetry is perfectly accounted for by means of (advanced and 
retarded) propagators, see for example~\cite{BENATTI} and also~\cite{LS2}.

The resonance states originally introduced by Gamow do not need the Hardy
axiom to be well defined. Moreover, as will be shown in a forthcoming
paper, the resonance states of the Bohm-Gadella theory, which are defined as
\begin{equation}
      |z_{\rm R}^-\rangle \equiv \frac{\rmi}{2\pi} 
                   \int_{-\infty}^{+\infty}\rmd E \, 
            \frac{|E^-\rangle}{E-z_{\rm R}} \, , 
     \label{Bresos}
\end{equation}
where $z_{\rm R}$ is the resonance energy, are not the same as Gamow's 
original state. In addition, the ``Gamow vectors''~(\ref{Bresos}) make use
of the whole real line of energies. Thus, they do not make 
physical sense, because they assign a physical meaning to the decay of a 
resonance state $|z_{\rm R}^-\rangle$ into a scattering state $|E^-\rangle$ 
of negative energy,
\begin{equation}  
       \langle ^-E|z_{\rm R}^-\rangle = \frac{\rmi}{2\pi} \frac{1}{E-z_{\rm R}} \, , 
\end{equation}
where $E$ belongs to the second sheet for $E<0$. By contrast, quantum
mechanical decay only occurs into energies of the 
physical spectrum. Once again, the negative energies are used inappropriately.

\section{Conclusions}
\label{sec:conclusions}

A thorough examination of the Bohm-Gadella theory has revealed its
inconsistency with standard quantum mechanics:
\begin{itemize}
   \item[$\bullet$] Contrary to the Hardy axiom, the wave functions 
$\varphi ^+(E)$ and $\psi ^-(E)$ are not Hardy functions.
   \item[$\bullet$]  The derivation of the preparation-registration arrow of 
time is flawed. 
    \item[$\bullet$] The resonance states of the Bohm-Gadella theory are not
the same as Gamow's original states.
    \item[$\bullet$] That $\varphi ^+(E)$ and $\psi ^-(E)$ are not 
Hardy functions seems to stem from the semiboundedness of the Hamiltonian, 
because Hardy functions seem more adequate to describe causality in systems 
where energy (frequency) integrations are taken over the whole real line.  
    \item[$\bullet$] The Bohm-Gadella theory does not rest on the same 
principles as quantum mechanics.
    \item[$\bullet$] The major achievements of the Bohm-Gadella theory,
namely time asymmetry and the rigorous construction of resonance states, can 
be achieved rigorously within standard quantum mechanics. 
\end{itemize}    
The Bohm-Gadella theory must therefore be abandoned. Not abandoning the 
Bohm-Gadella theory would force us ``{\it to bend the rules of standard 
quantum mechanics}'' 
(see the abstract of~\cite{PRA02}), which is unnecessary, because 
standard quantum mechanics is capable of describing scattering, decay and 
time asymmetry in a consistent manner.

\ack

This research was supported by MEC fellowship No.~SD2004-0003.

\end{document}